\documentclass[aps,pre,showpacs,showkeys,twocolumn,floatfix]{revtex4}

  \usepackage{graphicx,subfigure}
\usepackage{epsfig} 
\usepackage{epstopdf}
\usepackage{color}
\usepackage{amsfonts}
\usepackage{amssymb,amsmath}
 \usepackage{appendix}

\begin{document}

\title{
Symmetry breaking by heating in a continuous opinion model\\
}
 
\author{Celia Anteneodo$^{1,2}$}
\thanks{celia.fis@puc-rio.br}

\author{Nuno Crokidakis$^{3}$}
\thanks{nuno@if.uff.br}

\affiliation{
$^{1}$Departamento de F\'isica, PUC-Rio, Rio de Janeiro/RJ, Brazil \\
$^{2}$National Institute of Science and Technology for Complex Systems, Brazil\\
$^{3}$Instituto de F\'{\i}sica, Universidade Federal Fluminense, Niter\'oi/RJ, Brazil }

\date{\today}

\begin{abstract}
\noindent

We study the critical behavior of a continuous opinion model, driven by kinetic exchanges in a 
fully-connected population. 
Opinions range in the real interval $[-1,1]$, representing the different shades of opinions 
against and for an issue under debate. 
Individual's opinions evolve through pairwise interactions, 
with couplings that are typically positive,  but a fraction $p$ of negative ones is allowed. 
Moreover, a social temperature parameter $T$ controls the tendency of the individual 
responses towards neutrality.
Depending on  $p$ and $T$, different collective states emerge: symmetry broken (one side wins), 
symmetric (tie of opposite sides) and absorbing neutral (indecision wins). 
We find the critical points and exponents that characterize  the phase transitions between them.  
The symmetry breaking transition belongs to the usual Ising mean-field universality class, 
but the absorbing-phase transitions, with $\beta=0.5$, are out 
of the paradigmatic directed percolation  class.  
Moreover,  ordered phases can emerge by increasing social temperature.

\end{abstract}



\maketitle

\section{Introduction} 

Models of opinion formation  are useful to understand the requirements for the upraise of 
consensus and polarization, 
when a given issue is under debate~\cite{loreto_rmp,galam_book,sen_book,galam_review}.  
Although interactions between agents are different from those that govern physical systems,   
similar collective states and transitions between them can emerge~\cite{loreto_rmp,galam_book,sen_book,galam_review,pre_2010}. 

Opinions can be modeled by a discrete variable, to account for a restricted number of 
options~\cite{galam_review,sen,biswas,galam,javarone,javarone2,nuno_celia,xiong,pan,sznajd,sibona,biswasPRE2011,plurality,balakin,chatterjee}. 
For instance, the opinion $o_i$ of individual $i$ can take the values -1,0,1, 
corresponding to the unfavorable, neutral and favorable attitudes.  
The opinion can also be given by a continuous variable, for instance, in the real range $[-1,1]$, 
in order to represent the   possible shades of individual's attitudes  against or for the topic 
under discussion ~\cite{marlon,fan,hk,deffuant,deffuant3,lorenz,coda,jstat,lccc,biswas_conf,brugna}.
A simple rule for the evolution of opinions, be discrete or continuous, which has been considered 
previously~\cite{sen,lccc,biswas,nuno_celia,victor},   is given by 
\begin{equation} \label{eq0}
o_{i}(t+1) =   o_{i}(t)+\mu_{ij}\,o_{j}(t),  
\end{equation}
with $1\le i,j \le N$,  $j\neq i$,  where $N$ is the population size, and $\mu_{ij}$ are coupling coefficients 
that take into account the contribution of agent $j$ to 
the new opinion of agent $i$ at time $t+1$, given the opinions at the previous time $t$. 
Therefore, the coefficients $\mu_{ij}$ can be viewed as elements of a weighted adjacency matrix, 
where elements are null for non connected individuals, 
and non-null elements represent the strength of interactions. 
These coefficients are typically positive, meaning a positive influence, 
or agreement  but, because disagreement also exists in a population \cite{vaz}, 
a fraction $p$ of negative influences is allowed. 
Since the right-hand-side of Eq.~(\ref{eq0}) can exceed the extreme values $\pm 1$, 
an additional rule is required to forbid changes that exceed limiting values 
(or, equivalently, to re-inject the opinion  back  to the corresponding  end value). 
This additional rule introduces  nonlinearity into rule (\ref{eq0}). 

Kinetic equation (\ref{eq0}) has been studied in the literature for discrete~\cite{biswas,nuno_celia,victor} 
and continuous~\cite{biswas,jstat} opinion variables. 
In both cases, it produces a continuous phase transition, which resembles the ferromagnetic-paramagnetic one, 
when the  disorder parameter $p$ that controls the fraction of negative influences is varied. 

In this work, we modify Eq.~(\ref{eq0}) as follows
\begin{equation} \label{eq1}
o_{i}(t+1) =  \tanh   \Bigl(  [o_{i}(t)+\mu_{ij}\,o_{j}(t) ]/T \Bigr) ,
\end{equation}
where  $T$ is a positive real parameter that plays the role of a social temperature, 
a degree of randomness in the behavior of individuals \cite{kacperski,holyst,kacperski2,lalama}. 
The hyperbolic tangent naturally restricts the opinions to the interval $[-1,1]$, 
in contrast to rule (\ref{eq0}) that needs an additional prescription. 
In such case,  the nonlinearity is of the type linear by parts, while the nonlinearity in  Eq.~(\ref{eq1}) is smooth.
The lower the temperature $T$, the more each individual opinion  $o_i(t)$ will tend to adopt one of
the extreme values $\pm 1$. Meanwhile, for high temperature, the new opinion will tend to be 
more neutral. 
Therefore, $T$ plays a role different from the temperature in Metropolis dynamics, which promotes antialignement of 
interacting particles.   
As we will see, the dynamics governed by Eq.~(\ref{eq1}) recovers  discrete and continuous scenarios, but also introduces 
new features, in particular,   phase transitions to absorbing states. 


\section{Model}

Each agent $i$ has an opinion $o_{i}$ in the real range $[-1,1]$. 
Opinions tending to $o=\pm 1$ indicate extremist individuals, while  
  opinions $o\approx 0$ mean neutral or undecided ones. 
We will consider populations of  $N$  fully-connected individuals, situation 
which corresponds to a mean-field limit.  
Moreover, the couplings $\mu_{ij}$, which are either negative or positive 
with probabilities $p$ and $1-p$, respectively, 
are drawn from uniform distributions in the real intervals $[-1,0)$ and $[0,1]$, respectively.
Concerning the random nature of the coefficients $\mu_{ij}$, 
we  consider quenched (frozen) variables, 
as far as opinion formation supposedly occurs in a time scale much faster than the 
changes in agents' connections.

The initial state of the system is assumed to be fully disordered, that is, 
at the beginning of the dynamics, each individual has an opinion drawn 
from the uniform  distribution  in the range $[-1,1]$. 
At a given step  $t$, we  randomly choose two connected agents $i$ and $j$ and 
update opinion $o_i(t)$  according to Eq.~(\ref{eq1}). 
Notice that, as in Eq.~(\ref{eq0}),  Eq.~(\ref{eq1}) changes only the opinion of agent $i$ at step $t+1$. 
$N$ of such updates define the unit of time.


To characterize the coherence of the collective state of the population, we employ  

\begin{equation} \label{O}
O  =   \frac{1}{N}\left|\sum_{i=1}^{N} o_{i}\right|  ~. 
\end{equation}
Notice that this is a kind of order parameter 
that plays the role of  the ``magnetization per spin'' in  magnetic systems. 
It is sensitive to the unbalance between positive and negative opinions. 
A collective state with a significantly non-null value of $O$ means a symmetry-broken distribution of 
opinions.  Therefore,  the debate has a clear result, be extremist or moderate.  
A very small value of $O$ ($O\simeq 0$, within finite-size fluctuations) indicates a symmetric distribution of opinions, 
meaning that the debate will not have a winner position, but a tie. 
Finally, an exact value $O=0$, for finite size, indicates an absorbing state, where all agents share the neutral opinion $o=0$.

We compute $ \langle O \rangle $, where $\langle\, ...\, \rangle$ 
denotes average over random couplings and initial configurations. 
Furthermore, we calculate the fluctuations $V$  (or ``susceptibility'') of  parameter $O$, 
\begin{equation} \label{V}
V =  N\,(\langle O^{2}\rangle - \langle O \rangle^{2})  \,, 
\end{equation}
and the Binder cumulant $U$~\cite{binder}, defined as
\begin{equation} \label{U}
U   =   1 - \frac{\langle O^{4}\rangle}{3\,\langle O^{2}\rangle^{2}} \,.
\end{equation}

 \section{Results}
\label{results}

\begin{figure}[h]
\begin{center}
\includegraphics[width=0.45\textwidth,angle=0]{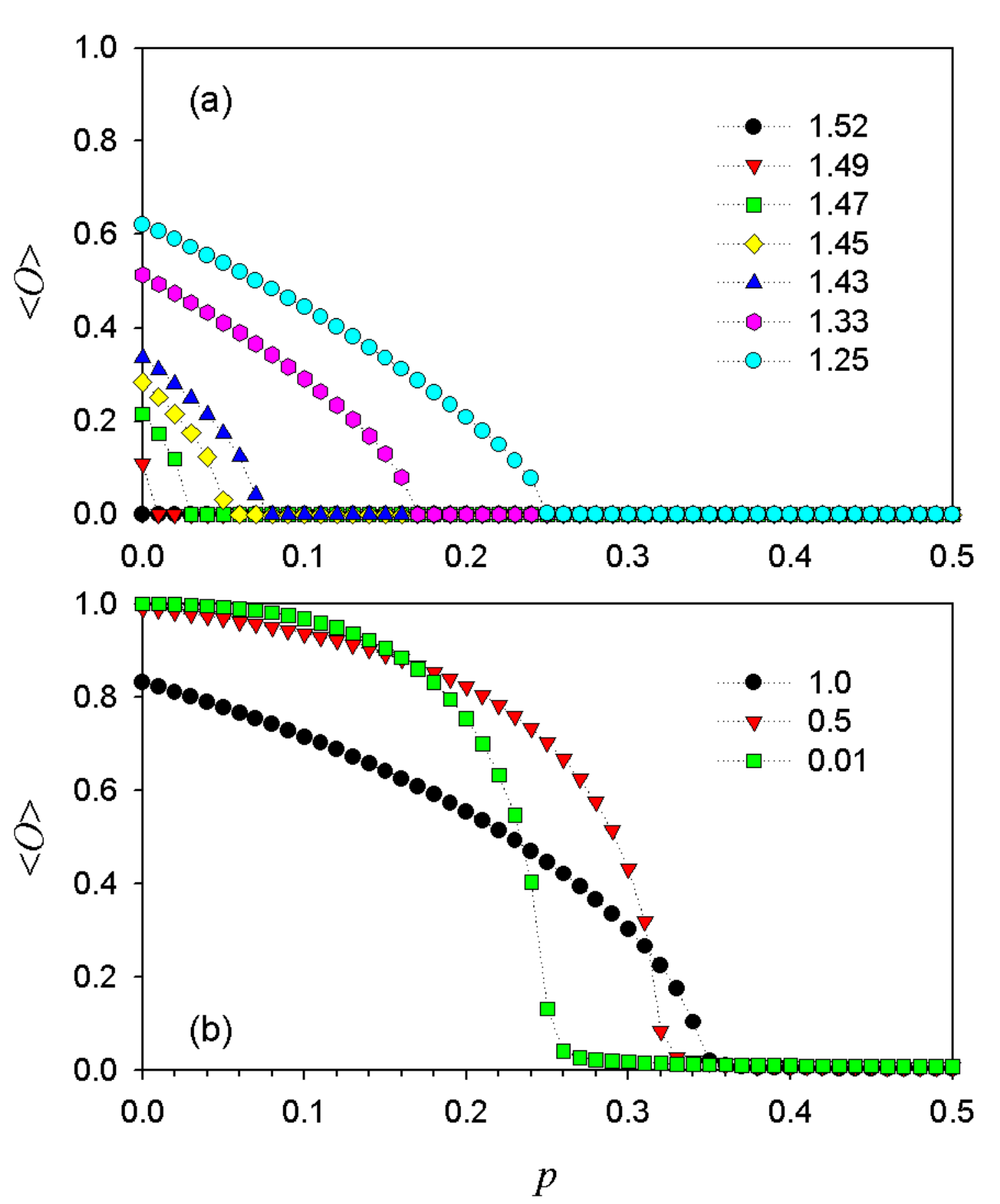}
\end{center}
\caption{Steady value of the order parameter  $\langle O \rangle$   versus $p$, 
for different values of $T$ indicated on the figure. 
(a) For high social temperature $T$, there is a transition from an asymmetric phase to 
an absorbing state where all opinions are neutral. 
(b) For low $T$, a transition from a symmetry-broken phase to a fluctuating symmetric one takes place. 
The population size is $N=10^{4}$ and data are averaged over $100$ simulations. 
 }
\label{fig:Oxp}
\end{figure}

The dependency of the steady value of the order parameter  $\langle O \rangle$, 
on the fraction of negative interactions $p$, is illustrated in Fig.~\ref{fig:Oxp}, 
for several values of the social temperature $T$, 
at finite population  size $N=10^4$.
For sufficiently high $T$ (Fig.~\ref{fig:Oxp}.(a)), 
the collective state  has broken symmetry (ferromagnetic-like) at low $p$, 
while above a critical value $p_c$,  
the order parameter $\langle O \rangle$ becomes exactly null  even for finite size. 
As stated above, this means that all the opinions became null, which corresponds to an absorbing state:  
once attained, it becomes frozen, because no further changes are possible~\cite{biswasPRE2011,biswas_conf,nuno_absorbing}.   
The transition between the asymmetric phase  and the absorbing state will be labeled I. 
Differently, for small $T$ (Fig.~\ref{fig:Oxp}.(b)), 
there is another kind of  transition (let us call it II), between the asymmetric phase and 
a fluctuating symmetric (paramagnetic-like) one, 
where the value of the parameter $\langle O \rangle$ is very small but vanishes only in the large size limit.
 This is the same kind of symmetry-breaking transition observed for Eq.~(\ref{eq0}).

\begin{figure}[h]
\begin{center}
\includegraphics[width=0.45\textwidth,angle=0]{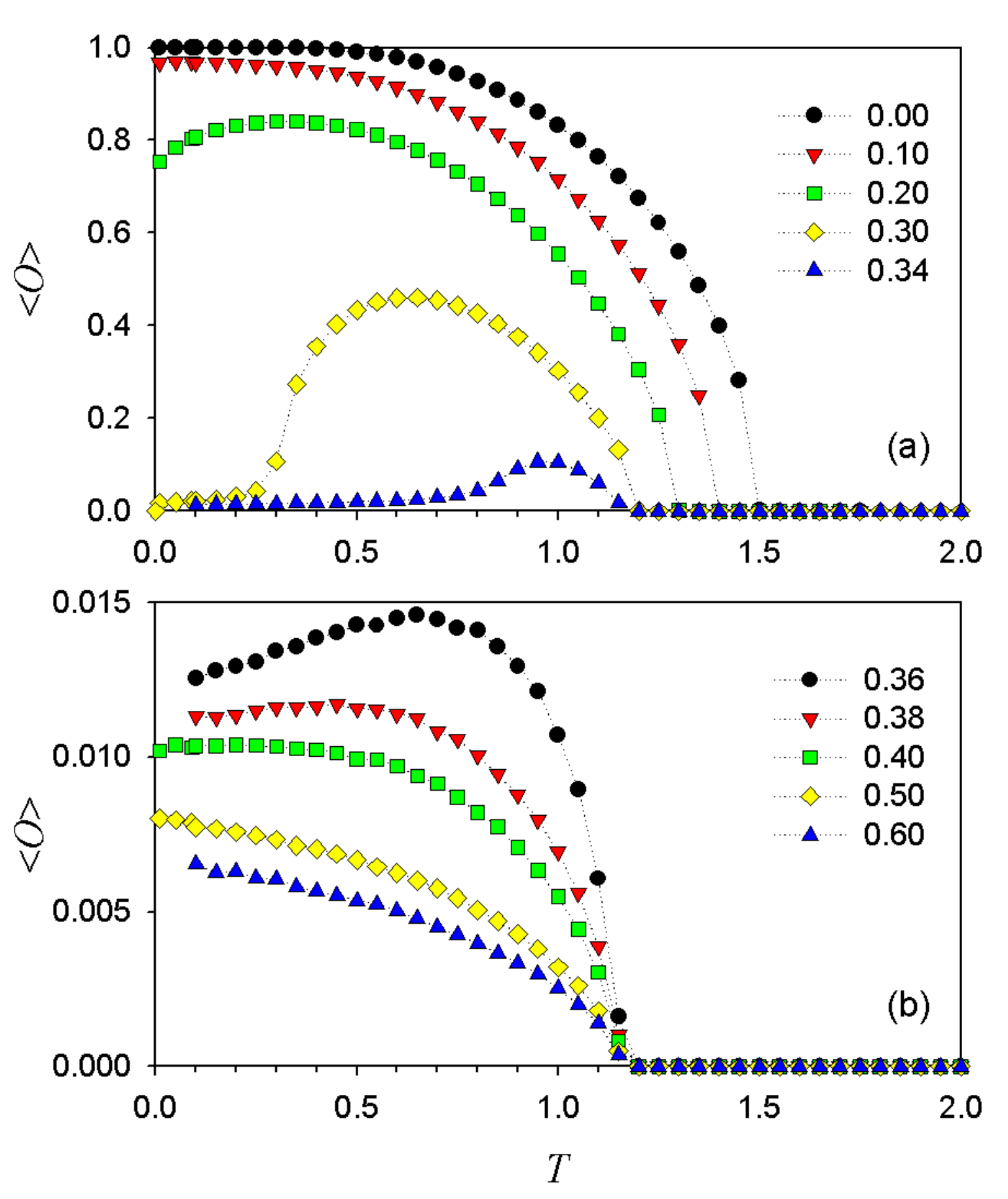}
\end{center}
\caption{Steady value of the order parameter $\langle O \rangle$ versus the social temperature $T$,  
for different values of $p$ indicated on the figure.  
Averages are computed over 100 samples, and $N=10^4$. 
 }
\label{fig:OxT}
\end{figure}

It is also interesting to watch $\langle O \rangle$ versus  $T$, 
for fixed values of $p$, as shown in Fig.~\ref{fig:OxT}. 
In Fig.~\ref{fig:OxT} (a), we observe transitions of the same type described above when $p$ was varied. 
A transition of type I occurs at a critical $T$, for low enough $p$. 
It is noticeable that the order parameter $\langle O \rangle$ first increases with $T$ for moderate values of $p$.
For larger $p$, besides the transition of type I,  a transition of type II also occurs, at a critical $T$ lower than the 
critical temperature of transition I. 
But, for still larger $p$ (Fig.~\ref{fig:OxT} (b)), a new kind of transition  appears (type III): 
 between a fluctuating symmetric phase and the absorbing state.

\begin{figure}[h]
\begin{center}
\includegraphics[width=0.5\textwidth,angle=0]{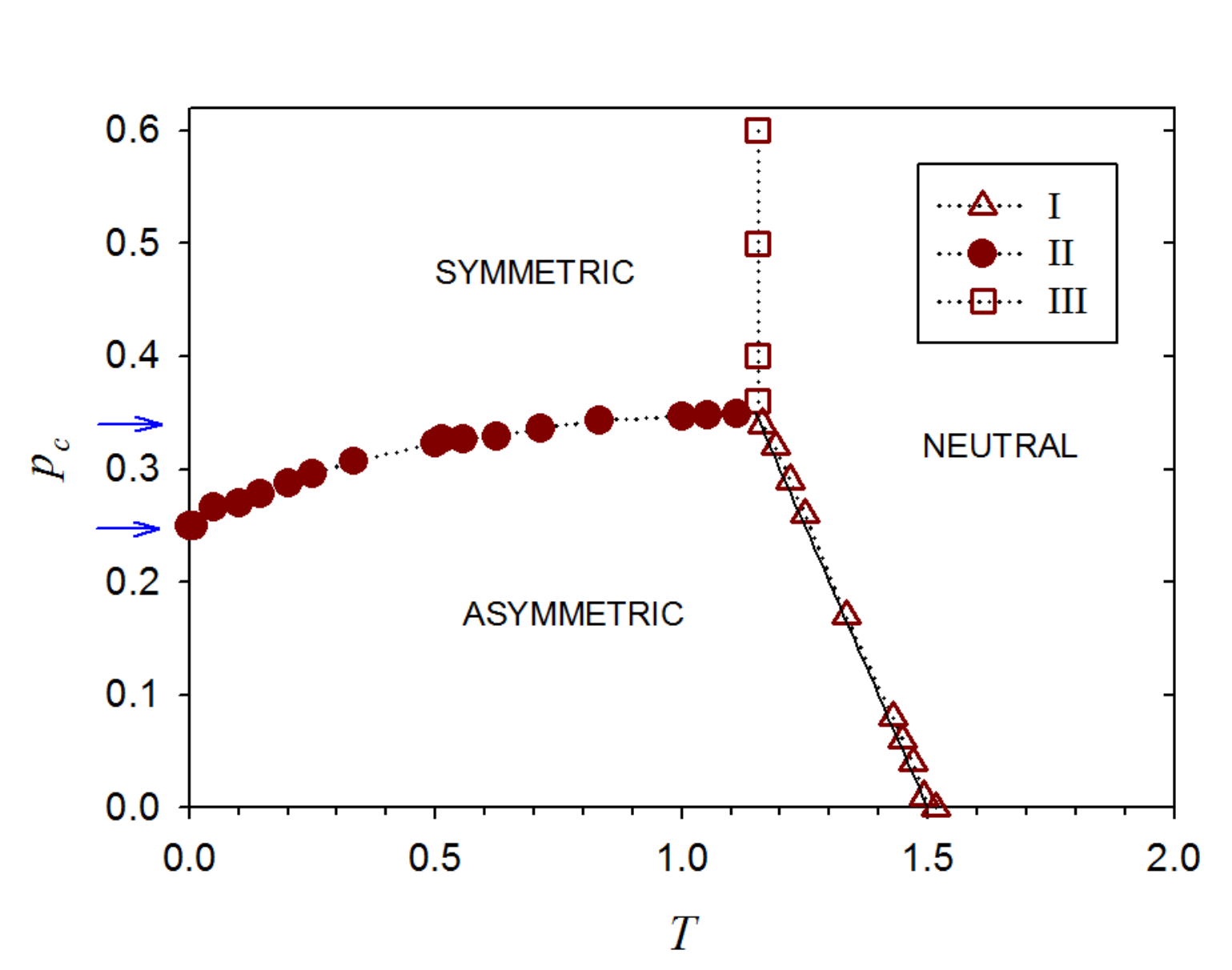}
\end{center}
\caption{Critical points $p_{c}$ as a function of $T$, for the transitions  
between symmetry-broken, symmetric and neutral (absorbing)  collective states. 
For more details about the distribution of opinions in each phase, see Fig.~\ref{fig:histos} 
in Appendix~\ref{sec:scaling}.
The arrows highlight  the values for the discrete ($p_c^{II}=0.25$)  and 
the continuous ($p_c^{II}\approx 0.34$) opinion models~\cite{biswas}.  
The full line is predicted by Eq.~(\ref{teo2}). 
}
\label{fig:pc_xbeta}
\end{figure}

The critical points between the different phases are summarized in Fig.~\ref{fig:pc_xbeta}. 
Moreover, to provide a detailed picture of each phase, the distributions of opinions are shown and 
discussed in Appendix~\ref{sec:histos}.

In order to characterize the phase transitions, we performed scaling analyses, 
that allowed to determine the critical points, shown in Fig.~\ref{fig:pc_xbeta}, as well as critical exponents. 
 
For the symmetry-breaking transition II, we considered as usual the plots of 
the quantities defined in Eqs. (\ref{O})-(\ref{U}) versus $p$, for fixed $T$ 
(as illustrated in Appendix~\ref{sec:scaling}). 

The critical points $p_{c}^{II}(T)$, estimated from the intersection of the Binder cumulant curves 
are shown by the full circles in Fig.~\ref{fig:pc_xbeta}. 
The value $p_c^{II}=0.25$, obtained previously for the discrete opinion 
model~\cite{sen,lccc,biswas,nuno_celia,victor}, 
is recovered in the limit of small social temperature. The result for continuous opinions in the range $[-1,1]$, $p_c^{II}\simeq 0.34$~\cite{biswas}  is achieved at  $T\approx 1$. 

The critical exponents were obtained based on the scaling form
\begin{equation} \label{fss}
\langle O^k \rangle \sim    N^{-k\beta/\nu} F_k ( [p-p_c] N^{1/\nu}  )~, 
\end{equation}
for $k=1,2,\ldots$, where $F_k$ are scaling functions. 
As expected due to the mean-field character of the system, 
we found the usual mean-field Ising critical exponents $\beta\approx 1/2$, 
and $\nu\approx 2$~like for Eq.~(\ref{eq0})~\cite{sen,lccc,biswas,nuno_celia,victor} (see Appendix~\ref{sec:scaling}).

In the  case of the absorbing-state transitions, we determined the critical points 
as the first value for which the order parameter is null within a given error.  
The critical values for transitions I and III are respectively shown 
by triangles and squares in Fig.~\ref{fig:pc_xbeta}. 
Moreover, we considered the dynamic scaling ansatz~\cite{dickman,hinrichsen}
\begin{equation} \label{dyn_fss}
\langle O(t)\rangle  \sim  t^{-\beta/\nu}F( \Delta \,t^{1/\nu}) \,,
\end{equation}
where $t$ is the simulation time, $\Delta$ represents either $|T-T_c|$ or  $|p-p_c|$,   
and $F$ a scaling function that behaves as follows. 
In the non-absorbing phase, $F(x)\sim x^\beta$ for large $x$,  such that  the order parameter becomes time 
independent in the long time limit, and $F(x)$  tends to a constant value in the opposite limit of vanishingly small $x$, 
such that it decays as $t^{-\beta/\nu}$ at the critical point. 
In the absorbing phase, $F(x)$  tends to zero exponentially fast. 
 
\begin{figure}[h!]
\begin{center}
\includegraphics[width=0.5\textwidth,angle=0]{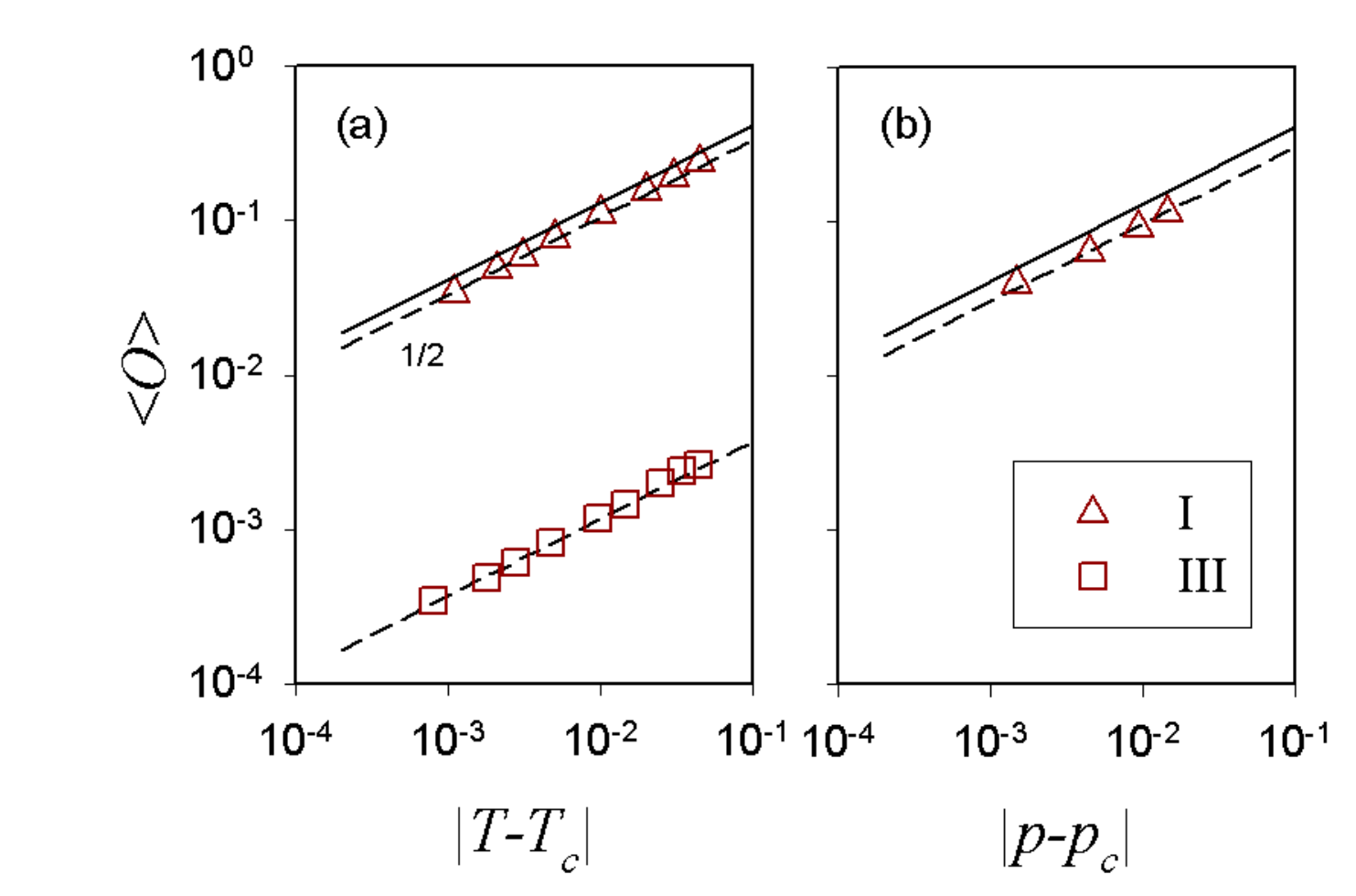}
\end{center}
\caption{Steady value of the order parameter as a function of $\Delta =|T-T_c|$ (a)  
 for $p=0.1$ ($T_c^I \approx 1.400$) and  $p=0.4$  ($T_c^{III}\approx 1.155$), and as a 
function of $\Delta =|p-p_c|$ (b) for $T=1.3$ ($p_c^{I}\approx 0.200$). 
In all cases the order parameter goes to zero at the critical point as $\langle O\rangle \sim \Delta^\beta$ 
with $\beta \approx 0.5$. The dashed lines with slope 0.5 are drawn for comparison. 
The solid lines are given by Eqs.~(\ref{teo3})-(\ref{teo4}).
In all cases, system size is $N=10^4$ and data were averaged over 100 realizations.
 }
\label{fig:otau}
\end{figure}

The exponent  $\beta$ is obtained  from the plots of $\langle O \rangle$ versus $\Delta$, where 
either $\Delta=|T-T_c|$ or $\Delta=|p-p_c|$ were considered. 
For all the absorbing-state transitions, the steady state value of the order parameter scales as 
$O\sim \Delta^\beta$,  with $\beta\approx 0.5$,   
as shown in  Fig.~\ref{fig:otau}.
Therefore these transitions are out of the directed percolation (DP) universality class, in contrast with 
results reported for  absorbing-state transitions of discrete opinion variants 
of Eq.~(\ref{eq0})~\cite{biswasPRE2011,nuno_absorbing}.

\begin{figure}[h!]
\begin{center}
\includegraphics[width=0.45\textwidth,angle=0]{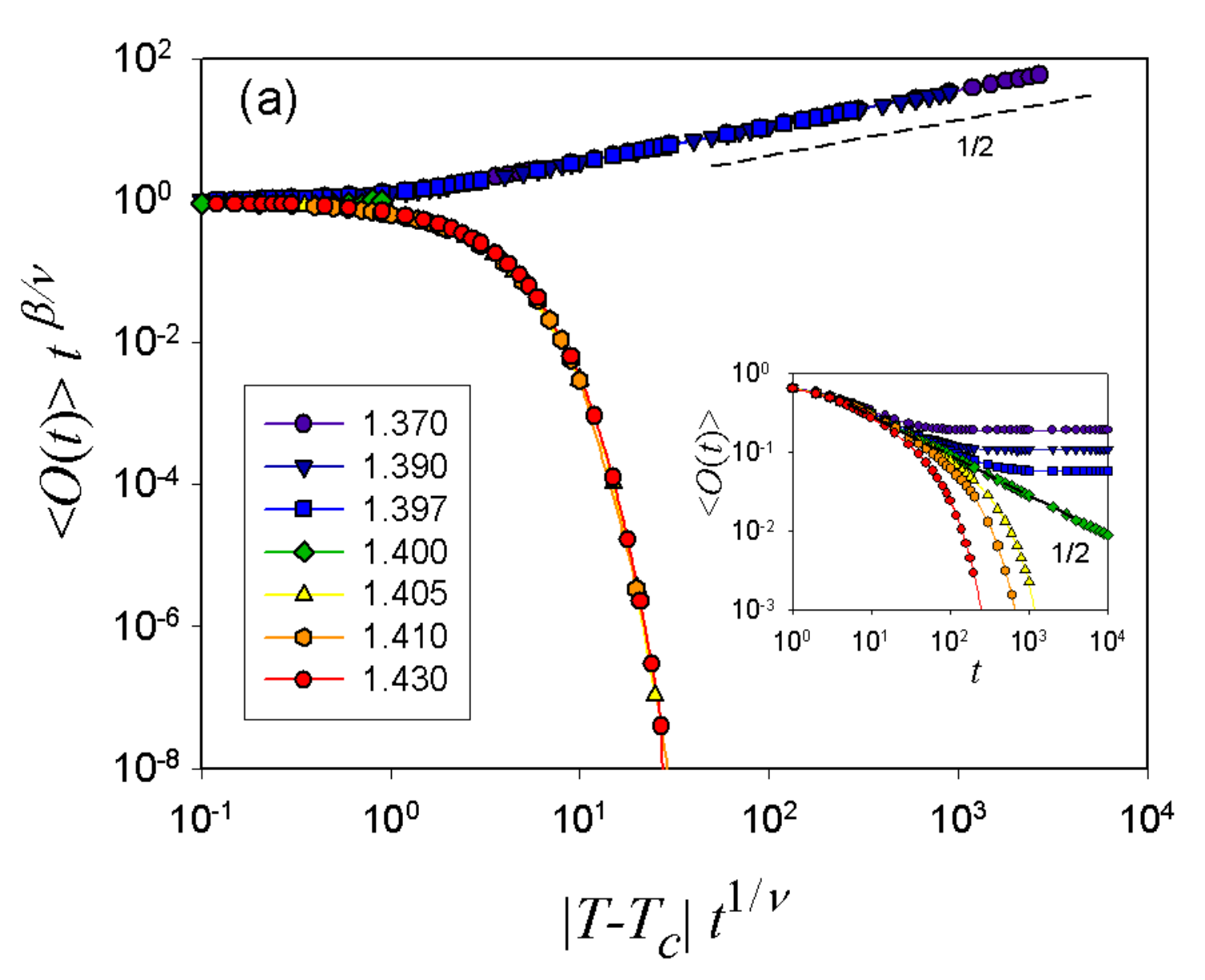}
\includegraphics[width=0.45\textwidth,angle=0]{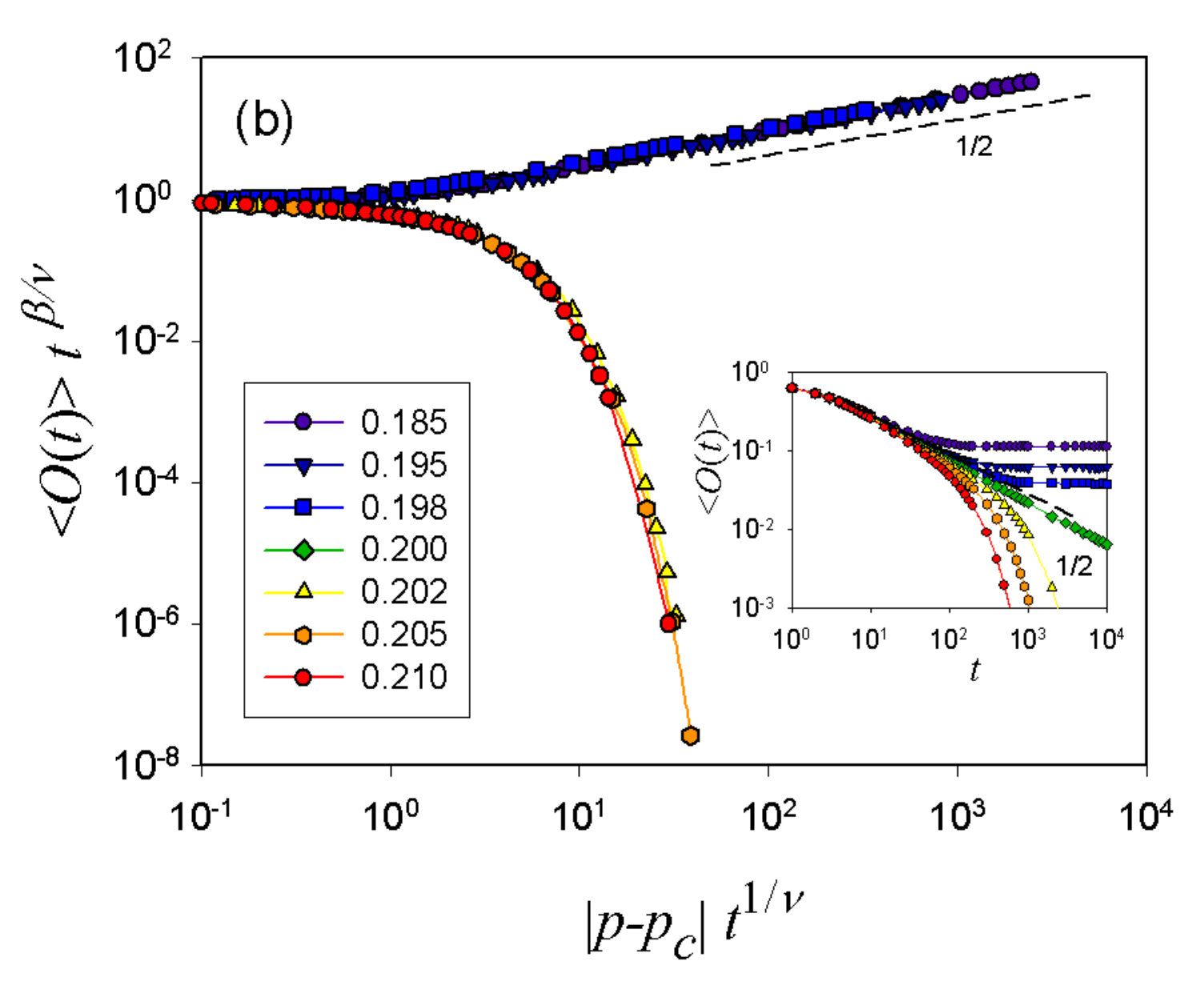}
\end{center}
\caption{Absorbing-phase transition I. Scaling plots of the order parameter, based 
on the scaling relation (\ref{dyn_fss}).
(a):  $\Delta=|T-T_c|$, for $p=0.1$ and different values of $T$ indicated on the figure, 
with data collapse obtained for $T_c^I\approx 1.400$, $\beta=0.5$  and $\nu=0.95$  
(b): $\Delta=|p-p_c|$, for $T=1.3$ and different values of $p$, with data 
collapse obtained for $p_c^I\approx 0.200$, $\beta=0.5$  and $\nu=0.98$.
The insets show the original non-scaled data. 
Data are averages over 100 samples, for population size $N=10^4$. 
The initial condition $o_i=1$ for all $i$ was used, 
although the final states are the same that for the random initial conditions 
used in other simulations.
}
\label{fig:dynFSS}
\end{figure}

Finally,  the value of $\nu$ ($\nu\equiv \nu_{||}$ in the literature) can be adjusted to produce the data collapse observed   
in  Fig.~\ref{fig:dynFSS}, yielding $\nu\approx 1$.  
Also for transition III (not shown), $\nu\approx 1$ was obtained.

The scaling of the order parameter at transition I can be understood heuristically as follows. 
If the distribution of opinions is narrow, that is, if $o_i(t)  \approx \langle O(t) \rangle $ for all $i$, 
when averaging both sides of Eq. ~(\ref{eq1})  over configurations and random couplings, we obtain  the map 
\begin{equation}  \label{teo0}
\langle O\rangle(t+1) =  \langle \tanh    \bigl(  \langle O\rangle(t)\, [ 1+  \mu_{ij} ]/T \bigr)\rangle \,, 
\end{equation}
where independence between couplings and opinions has been used.
For small argument of the hyperbolic tangent, at first order we have $\tanh x \approx x$, then 
Eq.~(\ref{teo0}) becomes
\begin{equation}  \label{teo1}
\langle O\rangle(t+1) \approx  \langle O\rangle(t)\, [ 1+  \langle\mu_{ij}\rangle ]/T  
= \langle O\rangle(t)\, (3/2-p)/T   \,.
\end{equation}
where the average  $\langle \mu_{ij} \rangle= (1-2p)/2$ comes from the distribution of the random couplings 
described at the beginning of Sec. II.
For $T>3/2-p$, the slope of the hyperbolic tangent at the origin is smaller than 1, 
then,  the  null fixed point of the map is stable, 
otherwise it becomes unstable and a stable non-null solution appears. 
Then, the critical value is
\begin{equation} \label{teo2}
p_c=3/2-T\,,
\end{equation} 
which provides the critical (solid) line represented in Fig.~\ref{fig:pc_xbeta}. 
At the following order, we have $\tanh x \approx  x -x^3/3$, 
then, from Eq.~(\ref{teo0}), using the result $\langle(1+\mu_{ij})^3\rangle =(15-14p)/4$, 
for the current statistics of couplings, 
its is straightforward to obtain the following 
expressions for the steady non-null solution, near the critical point:  
\begin{equation}  \label{teo3}
\langle O\rangle \approx   \Bigl[ 3(3-2p_c)^2/(15-14p_c) \Bigr]^{1/2} (p_c-p)^{1/2}
\end{equation} 
and
\begin{equation}  \label{teo4}
\langle O\rangle \approx    \Bigl[ 6T_c^2/(7T_c-3)  \Bigr]^{1/2}  (T_c-T)^{1/2} \,,
\end{equation}
(where $T_c=3/2-p$). This explains the exponent $\beta=0.5$ found numerically for transition I. 
This unusual value is a consequence of the cubic form of $\tanh$ near the origin,  
 while $\beta=1$ was reported for the linear by parts protocol~\cite{biswasPRE2011}.



\section{Final remarks}

We studied a model of opinion formation based on pairwise kinetic exchanges, in presence of a social temperature. 
A low temperature reinforces extreme attitudes while a high temperature softens extreme attitudes, 
promoting indifference or neutrality. 
Our target was to analyze the competition among such temperature and pairwise agent-agent 
interactions on the opinion formation process.

Concerning the critical behavior of the model, 
three different kinds of collective states and nonequilibrium transitions between them are observed, 
as summarized in Fig.~\ref{fig:pc_xbeta}.

At very high temperatures, the absorbing state where all opinions are neutral emerges for any $p$. 
However, depending on the fraction $p$ of negative couplings, different transitions are possible, when varying $T$. 

- For $p< 0.25$, only a transition of type I occurs, from the symmetry-broken phase, where one side dominates the debate,
 to the neutral absorbing state at high temperature.

- For $p\gtrsim 0.35 $, the distribution of opinions is symmetric for any temperature. 
But a transition of type III occurs  
from the fluctuating symmetric phase to the neutral absorbing  state  at high temperature. 
Since in the latter case, there is consensus of neutrality, it is more ordered 
than the fluctuating phase emerging at low $T$. 
To the best of our knowledge, it is the first time that such kind of  phase transition  is observed   
in continuous models of opinion formation.

- In the intermediate range  $0.25<p\lesssim 0.35$, curiously, the symmetric phase emerges 
at low temperature! and this phase suffers a  symmetry breaking transition  when the temperature 
overcomes a critical value. 
The critical point $p_c^{II}$ shifts to higher values with increasing $T$, indicating that the effects  
of $T$ compensate the disorder introduced by the negative influences, which occur with probability $p$. 
At higher $T$,  transition I  takes place  from the ordered phase to the neutral one. 
That is, for intermediate values of the fraction $p$ of negative couplings, 
the social temperature induces order, which is   another nontrivial result 
 that as far as we know has not been observed before in this kind of systems.  

As an extension of this work, one can consider the competition with other kinds of noises, like independence. It can be interesting to analyze the impact of such mechanisms on the critical behavior observed here.

\appendix


\begin{figure*}
\begin{center}
\includegraphics[width=\linewidth,angle=0]{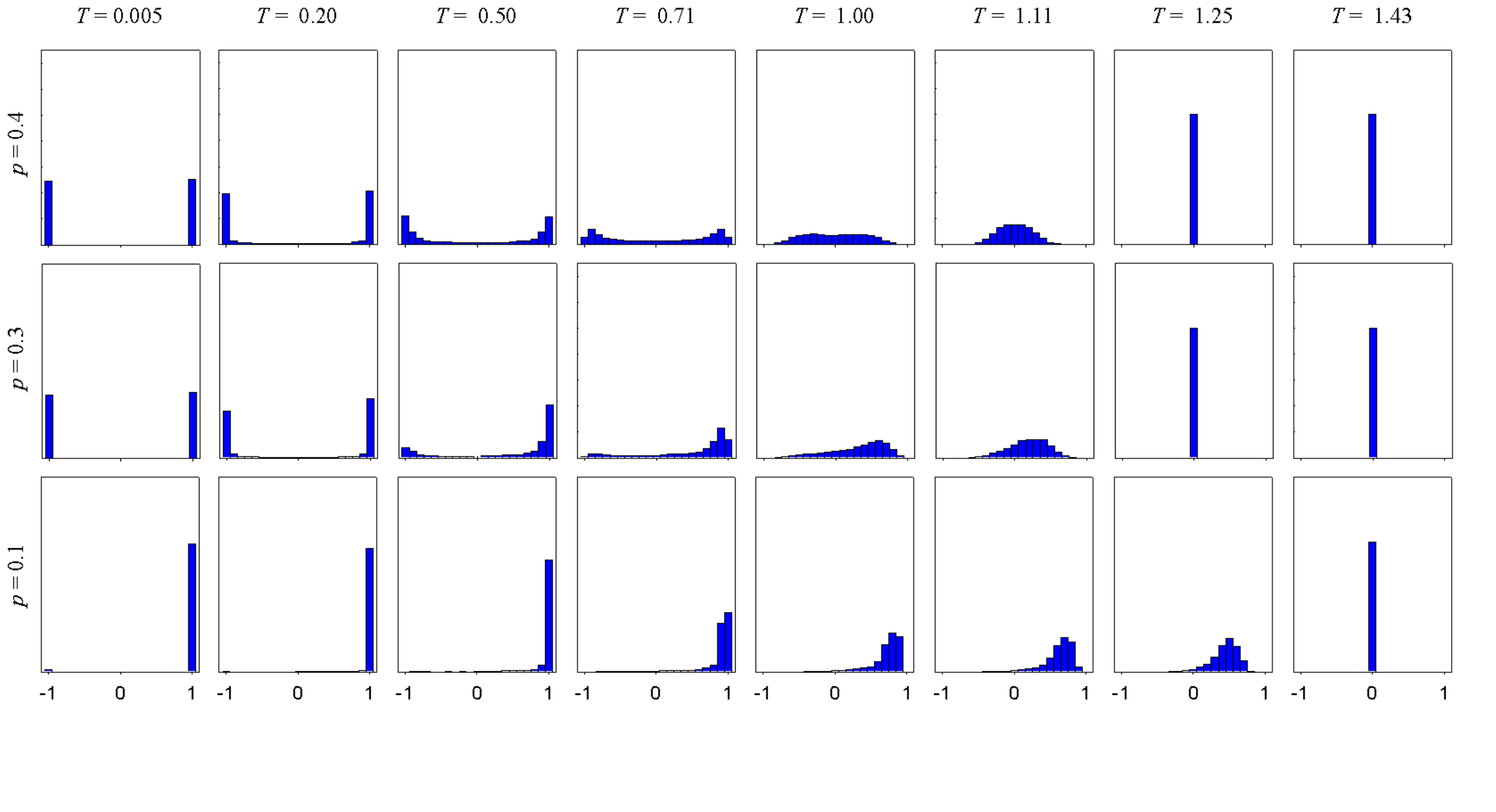}
\vspace*{-2cm}
\end{center}
\caption{Histograms of opinions at the steady state, for different values of $p$ and $T$. 
In all cases, for sufficiently high $T$, e.g. $T=1.43$, all opinions are neutral (hence, $\langle O\rangle=0$). 
As  $T$ decreases, we observe the following behaviors:
At $p=0.1$, when $T$ decreases below $T_c^I(0.1)\approx 1.400$,  
the distribution spreads asymmetrically and the peak shifts towards one of the extreme values. 
At $p=0.3$,  transition I, from neutral to biased, is observed at $T_c^I(0.3)\approx 1.20$, but in this case,  
as $T$ decreases, the distribution becomes bimodal and tends to recover symmetry, condensing at the extreme values.
At $p=0.4$, below $T_c^{III}(0.4)\approx 1.155$, the distributions spreads, always symmetrically in this case, 
becoming bimodal at very low $T$ such that, also in this case,  
only extreme opinions in the same proportion survive in the low $T$ limit. 
The population size is $N=10^{4}$ and data are averaged over $100$ simulations. 
 }
\label{fig:histos}
\end{figure*}

\section{Distributions of opinions}
\label{sec:histos}

The portrait of the  emergent collective states is depicted in Fig.~\ref{fig:histos} 
through the histograms of opinions in the steady state, 
for different values of $p$ and $T$.
It is interesting to compare these histograms with the phase-diagram of Fig.~\ref{fig:pc_xbeta}.  

At  high $T$ [e.g., last column ($T=1.43$), in Fig.~\ref{fig:histos}],  
we observe the absorbing state where all opinions become neutral ($o_i=0$ for all $i$, hence $\langle O\rangle =0$), 
independently of $p$. Let us analyze the changes that occur by decreasing $T$.

When crossing  the critical line I in Fig.~\ref{fig:pc_xbeta} from right to left, 
for instance for $p=0.1$ [$T_c^I(0.1)\approx1.400$],    
 the distribution, which is concentrated at $o=0$ when $T$ is high, first widens and suffers a bias (symmetry breaking). 
In this case, one side (either positive or negative)  dominates the debate, 
yielding a significantly non-null value of $\langle O \rangle$ in that phase. 
As $T$ decreases, the peak of the distribution is shifted towards one of the extreme values. 
But consensus (of one of the extreme opinions) is  attained only at $p=0$ (see also Fig.~\ref{fig:OxT}.a). 
 
Differently, when crossing the critical line III from right to left, for instance at $p=0.4$, 
the distribution widens but without losing its symmetry, giving rise to the disordered fluctuating phase, 
where positive and negative opinions are balanced. 
Moreover, the distribution which is always symmetric in this phase, 
becomes bimodal as $T$ decreases further, such that, for sufficiently low $T$, only the extreme values 
 survive, attaining a discrete portrait.  
However, for discrete opinions evolving according to Eq.~(\ref{eq0}), the three opinions (-1,0,1) coexist in 
the disordered phase~\cite{sen,biswas,nuno_celia,victor}.

For intermediate values of $p$, in the range [$0.25,\approx 0.35$], e.g. $p=0.3$, 
besides transition I, also transition II occurs  at lower temperature, $T_c^{II}(0.3)\approx 0.28$. 
Below that critical value, symmetry tends to be recovered ending, also in this case, in a balanced distribution of 
extreme values, at very low $T$.

Crossing line II,   by decreasing $p$, is associated to  emergence of bias (symmetry breaking), 
both in the unimodal and bimodal cases.

It is noteworthy that,  
when crossing the critical line II (from left to right in Fig.~\ref{fig:pc_xbeta}), 
by increasing $T$, there is a transition 
from  the disordered phase at low temperature  
to the ordered one at higher temperature (symmetry breaking by heating). 
  
Moreover, when crossing line III, by increasing $T$, 
there is a passage from the fluctuating symmetric phase to the  absorbing one. 
Therefore a more ordered state emerges by increasing $T$ also in this case.

We observe that, according to the model,  
for increasing disagreement among individuals (increasing $p$), 
both (positive and negative) sides coexist, 
and  the population adopts moderate opinions for a wide range of social temperatures.  
For rising temperature, there is predominance of neutral individuals, 
i.e., opinions distribute symmetrically  around zero, becoming totally neutral at high $T$. 
On the opposite case, of small $p$ 
(high agreement among individuals), extremists of one of the sides predominate   
whereas the other side tends to disappear  for rising social temperature. 
But further rising temperature makes individuals progressively more moderate until they become  
totally neutral also in this case.


\section{Finite-size scaling analysis of transition II}
\label{sec:scaling}

\begin{figure}[b!]
\begin{center}
\includegraphics[width=0.45\textwidth,angle=0]{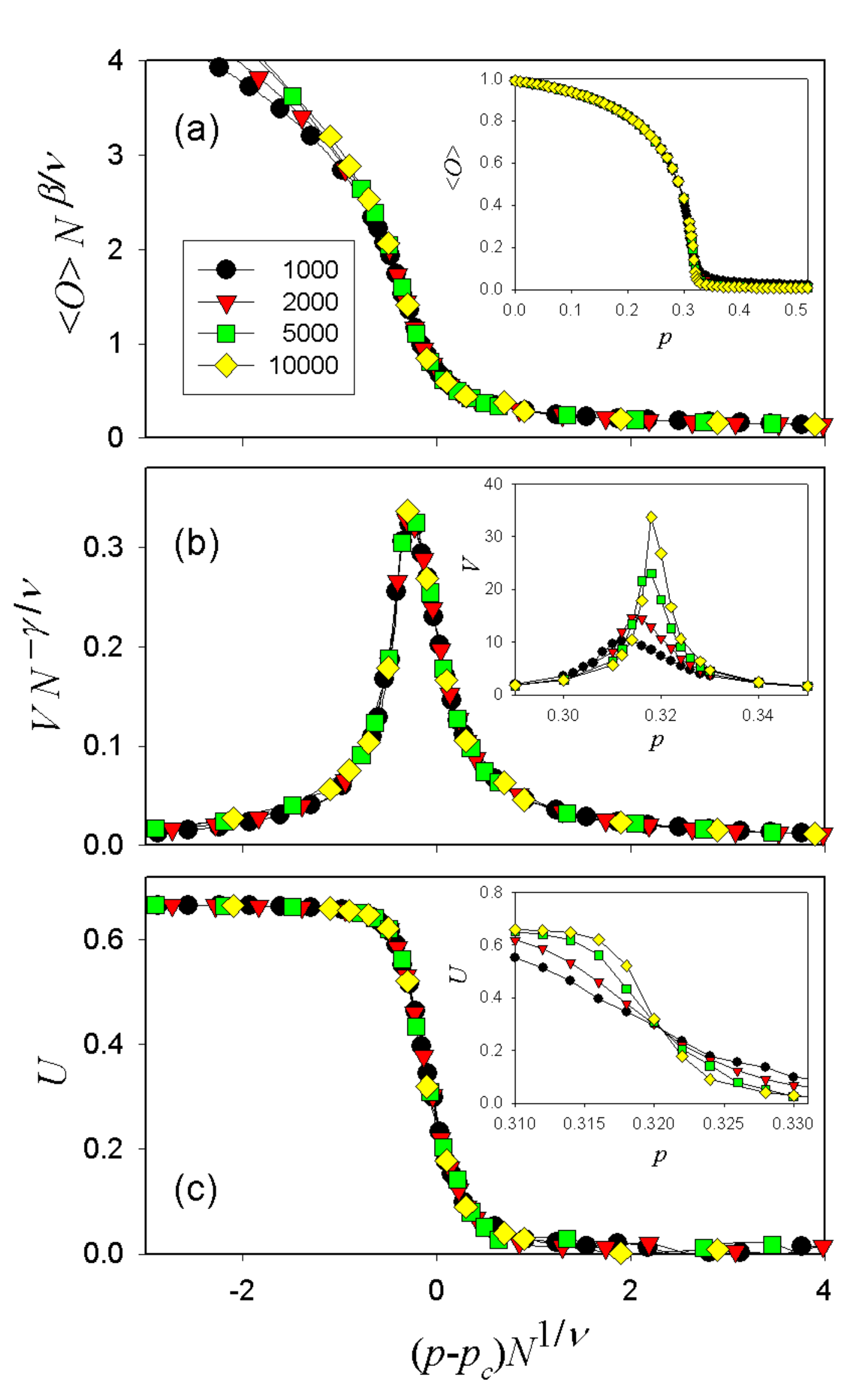}
\end{center}
\caption{Transition II. Finite-size scaling analysis of the quantities defined in Eqs. (\ref{O})-(\ref{U}), based 
on the scaling relation (\ref{fss}),  for $T=0.5$. 
The values of $N$ are indicated on the figure.
Data collapse was obtained for $p_c^{II}\approx 0.321$, $\beta=0.5$, $\nu=2$ and $\gamma= \nu-2\beta = 1.0$.
The insets show the original non-scaled data. 
}
\label{fig:FSS}
\end{figure}

For transition II (symmetry breaking), we analyzed the quantities defined in 
Eqs.~(\ref{O})-(\ref{U}), 
based on the scaling relation~(\ref{fss})~\cite{nuno_celia,victor}. 
Once obtained the critical value from the intersection of the Binder cumulant curves, we obtained $\nu$ and 
$\beta$ to produce data collapse. 
In the  case $T=0.5$,  shown in Fig.~\ref{fig:FSS}, 
data collapse was obtained for $p_c^{II}\approx0.321$, $\beta=0.5$, $\nu=2$, hence $\gamma= \nu -2 \beta =1.0$
This transition belongs to the mean-field Ising universality class.

\section*{Acknowledgments}

The authors acknowledge financial support from the Brazilian funding agencies CNPq and FAPERJ.

\end{document}